\documentclass{article}
\usepackage{spconf,amsmath,graphicx,hyperref}
\usepackage{algorithmic}
\usepackage{textcomp}
\usepackage{xcolor}
\usepackage{tcolorbox}
\usepackage{verbatim}
\usepackage{booktabs}
\usepackage{listings}
\usepackage{etoolbox}
\usepackage{tabularx}

\let\oldbibliography\thebibliography
\renewcommand{\thebibliography}[1]{%
  \oldbibliography{#1}%
  \setlength{\itemsep}{0pt}%
  \setlength{\parsep}{0pt}%
}

\title{Automatic Timestamping of Therapy Sessions \\Using Audio-Language Models - When and How Long Did Therapy Happen?}
\title{When and How Long Did Therapy Happen? \\
Soft-Supervising Temporal Localization Using Audio-Language Models}
\title{Fine-Tuning Large Audio-Language Models with LoRA for Precise Temporal Localization of Prolonged Exposure Therapy Elements}

\name{%
  \begin{tabular}{@{}c@{}}
    Suhas BN\textsuperscript{1},
    Andrew M.\ Sherrill\textsuperscript{2},
    Jyoti Alaparthi\textsuperscript{2},
    Dominik Mattioli\textsuperscript{1} \\
    Rosa I.\ Arriaga\textsuperscript{3},
    Chris W.\ Wiese\textsuperscript{4},
    Saeed Abdullah\textsuperscript{1}\thanks{This work is supported by the National Science Foundation (NSF) under Grant No.\ 2326144. Any opinions, findings, and conclusions or recommendations expressed in this work are those of the author(s) and do not necessarily reflect the views of the NSF.}
  \end{tabular}%
}
\address{%
\textsuperscript{1}College of Information Sciences \& Technology, The Pennsylvania State University, USA\\ \textsuperscript{2}Department of Psychiatry \& Behavioral Sciences, Emory University, USA\\ \textsuperscript{3}School of Interactive Computing, Georgia Institute of Technology, USA \\ \textsuperscript{4}School of Psychology, Georgia Institute of Technology, USA%
}
\begin{document}
\ninept
\maketitle

\begin{abstract}
Prolonged Exposure (PE) therapy is an effective treatment for post-traumatic stress disorder (PTSD), but evaluating therapist fidelity remains labor-intensive due to the need for manual review of session recordings. We present a method for the automatic temporal localization of key PE fidelity elements, identifying their start and stop times, directly from session audio and transcripts. Our approach fine-tunes a large pre-trained audio-language model, Qwen2-Audio, using Low-Rank Adaptation (LoRA) to process focused 30-second windows of audio-transcript input. Fidelity labels for three core protocol phases, therapist orientation (P1), imaginal exposure (P2), and post-imaginal processing (P3), are generated via LLM-based prompting and verified by trained raters. The model is trained to predict normalized boundary offsets using soft supervision guided by task-specific prompts. On a dataset of 308 real PE sessions, our best configuration (LoRA rank 8, 30s windows) achieves a mean absolute error (MAE) of 5.3s across tasks, within typical rater tolerance for timestamp review, enabling practical fidelity QC. We further analyze the effects of window size and LoRA rank, highlighting the importance of context granularity and model adaptation. This work introduces a privacy-preserving, scalable framework for fidelity tracking in PE therapy, with potential to support clinician training, supervision, and quality assurance.
\end{abstract}
%
%
\section{Introduction}
\vspace{-0.2cm}
Post-Traumatic Stress Disorder (PTSD) is linked to multiple co-morbidities and can severely impact quality of life. Recovery is possible through treatments like Prolonged Exposure (PE) therapy, a gold-standard, evidence-based approach ~\cite{foa2007prolonged}. However, trained PE clinicians are in short supply. Therapist fidelity, adherence to PE’s core components, is essential for effective outcomes~\cite{foa2013challenges}, but assessing fidelity typically requires manual review of full sessions, making it time-consuming and hard to scale. While prior work has explored automated evaluation using speech and language processing~\cite{6chen2021fusion,5li_bert2022,lin2023neural,lin2024compass}, most approaches focus on utterance-level classification rather than precise temporal localization of extended therapeutic phases, an essential requirement for PE fidelity assessment. Furthermore, the creation of large, finely annotated datasets remains a major bottleneck. This bottleneck effectively restricts high-quality care to a fraction of patients; thus, automating fidelity is not merely a technical convenience but an ethical imperative for healthcare equity. This paper introduces a framework for second-level temporal localization of key PE therapy fidelity phases, specifically~\cite{sherrill2021creating, Foa2019Prolonged}: therapist orientation to imaginal exposure (P1), duration of imaginal exposure (P2), and post-imaginal processing (P3). Our approach leverages Qwen2 Audio~\cite{chu2024qwen2}, a model that jointly processes audio and transcript inputs, fine-tuned using Low-Rank Adaptation (LoRA)~\cite{hu2021lora} and quantization techniques (QLoRA)~\cite{dettmers2023qlora} for computational efficiency. By distilling heavy server-side reasoning into a quantized local model, we ensure that patient audio, potentially the most sensitive biometric data, never leaves the secure inference environment, addressing a primary barrier to clinical AI adoption. To address annotation challenges, we employ a soft-supervision pipeline in which initial timestamps are proposed by a large language model (LLM) using session transcripts and subsequently verified by trained raters. We distill these high-quality annotations into a locally running model, avoiding external API calls and enabling privacy-preserving deployment in clinics or smartphones~\cite{bn2022privacy, suhas2023differential}. The model is trained on fixed-duration audio-transcript windows paired with task-specific prompts (e.g., \textit{``When did the Imaginal Exposure start?''}) to predict the normalized offset of the target event. Accurate temporal resolution is therefore critical, as the PE manual specifies that the three major components of the session, the initial check-in (P1), the imaginal exposure (P2), and the processing of the imaginal exposure (P3), should last approximately 10, 50, and 10 mins, respectively~\cite{sherrill2021creating, Foa2019Prolonged}. By automating these specific timestamps, we enable a ``human-in-the-loop'' workflow where supervisors need only review flagged segments rather than full sessions, potentially reducing the fidelity assessment burden by orders of magnitude.

Our contributions are:

\begin{enumerate}
    \item Proposed a method for fine-tuning Qwen2-Audio with LoRA for multimodal temporal localization in PE therapy.
    \item Introduced a novel soft-supervision strategy that combines LLM-based annotation with rater verification.
    \item Analyzed the impact of window duration and LoRA configuration choices.
    \item Demonstrated the approach on real-world PE therapy data.
    \item Presented a continuous temporal regression framework for fidelity tracking, building on advances in synthetic PE datasets \cite{bn2025thousand}, evaluation frameworks \cite{bn2025howreal}, and empathy modeling \cite{bn2025pursuit}, providing a foundation for scalable, fine-grained clinical assessment.
\end{enumerate}

To our knowledge, this is among the first efforts to fine-tune audio-language models for precise temporal localization of psychotherapy fidelity metrics. The remainder of this paper is organized as follows: Section ~\ref{sec:related_work} reviews related work, Section~\ref{sec:dataset} describes the dataset and annotation, Section~\ref{sec:methodology} details our methodology, Section~\ref{sec:experimental_setup} presents experimental setup, Section~\ref{sec:results_discussion} discusses the results, and Section~\ref{sec:conclusion} concludes.

\section{RELATED WORK}
\vspace{-0.2cm}
\label{sec:related_work}
Automated psychotherapy analysis scales feedback, enhances clinician training, and improves outcomes. We fine-tune pre-trained audio-language models on multimodal inputs to temporally localize fidelity metrics, using soft-supervised LLM-generated timestamps verified for clinical accuracy.

\vspace{-0.2cm}
\subsection{Automated Evaluation of Psychotherapy}
\vspace{-0.2cm}
Prior work has automated psychotherapy evaluation using NLP and speech processing. For session-level assessment, Chen et al. \cite{6chen2021fusion} developed a CBT quality pipeline using linguistic features, Flemotomos et al. \cite{flemotomos_automated2021} created a platform for MI skill evaluation through audio-based MISC coding, and Lin et al. \cite{lin2024compass} mapped dialogue to Working Alliance metrics. Li et al. \cite{5li_bert2022} classified client utterances in text counseling with BERT, while Lin et al. \cite{lin2023neural} applied topic modeling to understand session content. For specific behavior detection, researchers employed multimodal techniques with BERT, VGGish, and HuBERT for MI codes and empathy assessment \cite{tavabi2020multimodal, tran2023multimodal}; LSTMs with attention for addiction counseling behavior coding \cite{gibson2017attention, singla2018using, chen2019improving}; MEMMs for counselor reflection detection \cite{can2016sounds}; language synchrony for empathy \cite{lord2015more}; behavior trajectories in couples therapy \cite{tseng2017approaching}; and LLM embeddings for depression outcome classification \cite{xin2024using}. Prosodic analysis examined therapist speaking style \cite{tao2022characterizing}, empathy modeling \cite{xiao2014modeling}, and ChangeTalk valence prediction \cite{gupta2014predicting}. These studies primarily address utterance-level classification or segment-level scoring rather than precise temporal localization of broader fidelity metrics. 

\vspace{-0.2cm}
\subsection{Temporal Event Localization in Audio and Speech}
\vspace{-0.2cm}
Wu et al.'s FLAM \cite{wu2025flam} enables frame-wise open-vocabulary sound event detection through global and frame-level embeddings, while Mishra et al.'s SING \cite{mishra2025spatial} processes speech with Whisper and LLaMA to determine speaker identity and location. Both offer temporal localization capabilities but target different domains than psychotherapy fidelity metrics. 

\vspace{-0.2cm}
\subsection{Large Pre-trained Audio-Language Models}
\vspace{-0.2cm}
Recent audio-language models include Krishna D N's \cite{krishna2021using} cross-modal attention between Wav2Vec2.0 and BERT for emotion recognition, FLAM \cite{wu2025flam} and SING's \cite{mishra2025spatial} combination of speech understanding with LLMs, and Li et al.'s \cite{li2024audio} CLAP models associating audio and text. Our approach uniquely applies these models to localize psychotherapy fidelity metrics through specialized annotation and prompting. 

\vspace{-0.2cm}
\subsection{Multimodal Prompting \& Few-Shot Learning}
\vspace{-0.2cm}
Prior multimodal prompting approaches include Tsimpoukelli et al.'s ``Frozen'' \cite{tsimpoukelli2021multimodal} using images as visual prefixes, Li et al.'s PT-Text \cite{li2024audio} for ``audio-free'' CLAP tuning, text instruction prompts with LLMs in SING \cite{mishra2025spatial}, LLM prompting for interpretation \cite{lin2024compass}, and prompt engineering for data generation \cite{wu2024callm}. Our approach directly guides model attention during temporal fidelity metric localization.

\vspace{-0.2cm}
\subsection{Annotation and Supervision Strategies}
\vspace{-0.2cm}
We adopt a soft-supervision pipeline where LLM-proposed timestamps are verified by experts, offering a middle ground between intensive manual annotation in psychotherapy research \cite{6chen2021fusion, flemotomos_automated2021, wu2022anno} and fully automated approaches like FLAM's synthetic data generation \cite{wu2025flam} or CALLM's LLM-based mock profile augmentation \cite{wu2024callm}, balancing reduced expert burden with clinical validity.

\begin{figure}
    \centering
    \includegraphics[width=0.75\linewidth]{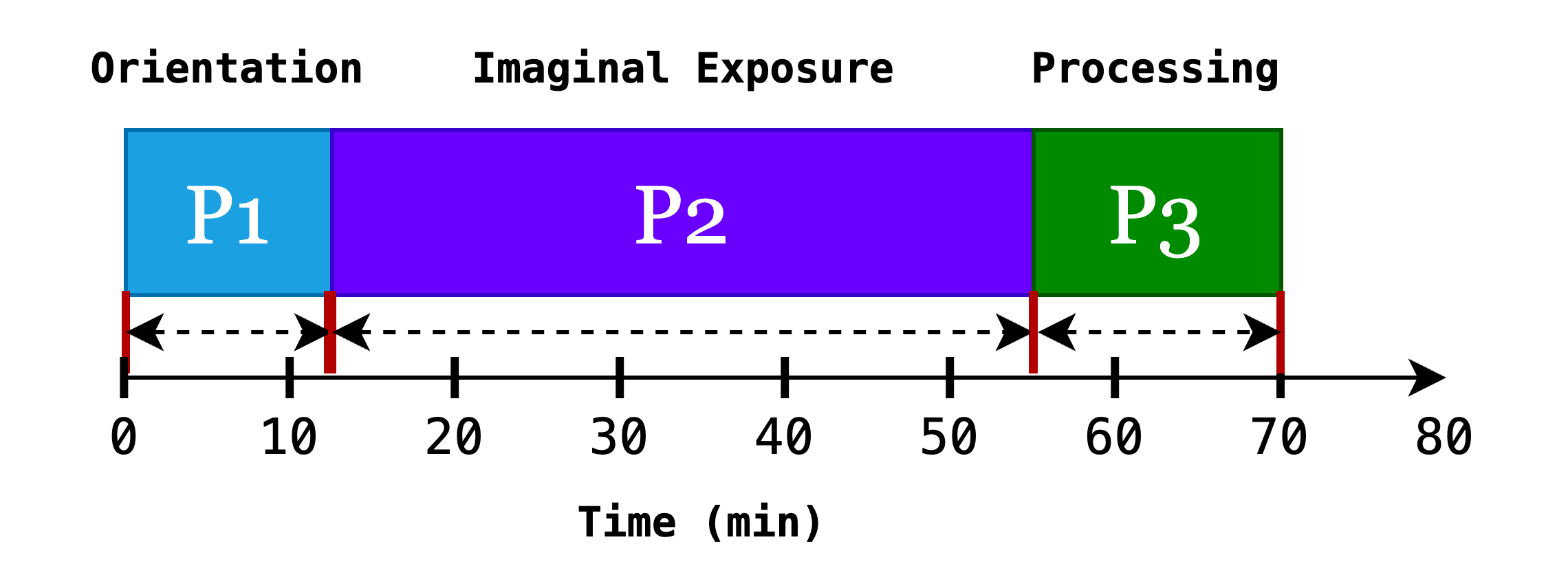}
    \caption{Timeline of a 70-minute PE therapy session showing typical segments for three fidelity metrics: P1 (orientation to imaginal exposure), P2 (imaginal exposure), and P3 (processing). Durations vary across sessions; this structure represents a common clinical pattern. All durations are in minutes.}
    \label{fig:session_split_up}
\end{figure}

\section{DATASET}
\label{sec:dataset}
\vspace{-0.2cm}
We use 318 PE therapy sessions (Table~\ref{tab:audio_stats}) recorded at Emory University, 60-90 minutes each. All data were handled in IRB-approved environments; no PII left secure storage. We annotated three protocol phases: \textit{orientation} (P1), \textit{imaginal exposure} (P2), and \textit{processing} (P3), which typically occur sequentially (Figure~\ref{fig:session_split_up}), motivating timestamp prediction. To reduce manual labeling, a zero-shot LLM extracted protocol presence and start/stop times from transcripts (JSON). For scalable validation, trained raters reviewed 10\% (30 sessions), yielding 94.4\% timestamp accuracy (within 5-10 s) and 97.7\% label accuracy. Audio was downsampled from 44.1-48 kHz to 16 kHz, normalized, and saved as WAV; transcripts with sentence-level timestamps and speaker labels were generated using Amazon HealthScribe. After excluding sessions over 1.5 hours, sessions with alignment issues, or those lacking valid labels, 308 recordings remained. We split the data into 216/45/47 train/validation/test sessions with balanced phase coverage.


\begin{table}[t!]
\centering
\scriptsize
\caption{Summary statistics of the audio dataset. Durations are shown in HH:MM:SS format, seconds in paranthesis.}
\label{tab:audio_stats}
\vspace{0.25cm}
\begin{tabular}{l r}
\toprule
\textbf{Statistic} & \textbf{Value} \\
\midrule
Number of sessions & 308 \\
Total duration & 337.76 hours \\
Min duration & 00:19:31 \hfill (1170.75 s) \\
Max duration & 01:29:54 \hfill (5393.86 s) \\
Average duration & 01:05:48 \hfill (3947.84 s) \\
Std. deviation & 00:15:36 \hfill (935.53 s) \\
\bottomrule
\end{tabular}
\end{table}

\begin{table}[t]
\centering
\scriptsize
\caption{Example LLM-generated annotations with timestamps (in sec) and ratings for PE therapy phases.}
\vspace{0.25cm}
\label{tab:llm_annotation}
\begin{tabularx}{\linewidth}{c X r r c}
\toprule
\textbf{ID} & \textbf{Phase Description} & \textbf{Start (s)} & \textbf{Stop (s)} & \textbf{Present} \\
\midrule
1  & Therapist orients the client to the planned imaginal exposure. & 4.17 & 37.27 & Yes \\
2 & Imaginal lasted about 30-45 minutes (or about 15 for final imaginal). & 105.83 & 2123.92 & Yes \\
3 & Therapist processes the imaginal exposure with the client. & 2123.92 & 3023.83 & Yes \\
\bottomrule
\end{tabularx}
\end{table}

\begin{figure}
    \centering
    \includegraphics[width=\linewidth, trim=0 30 0 0,
  clip]{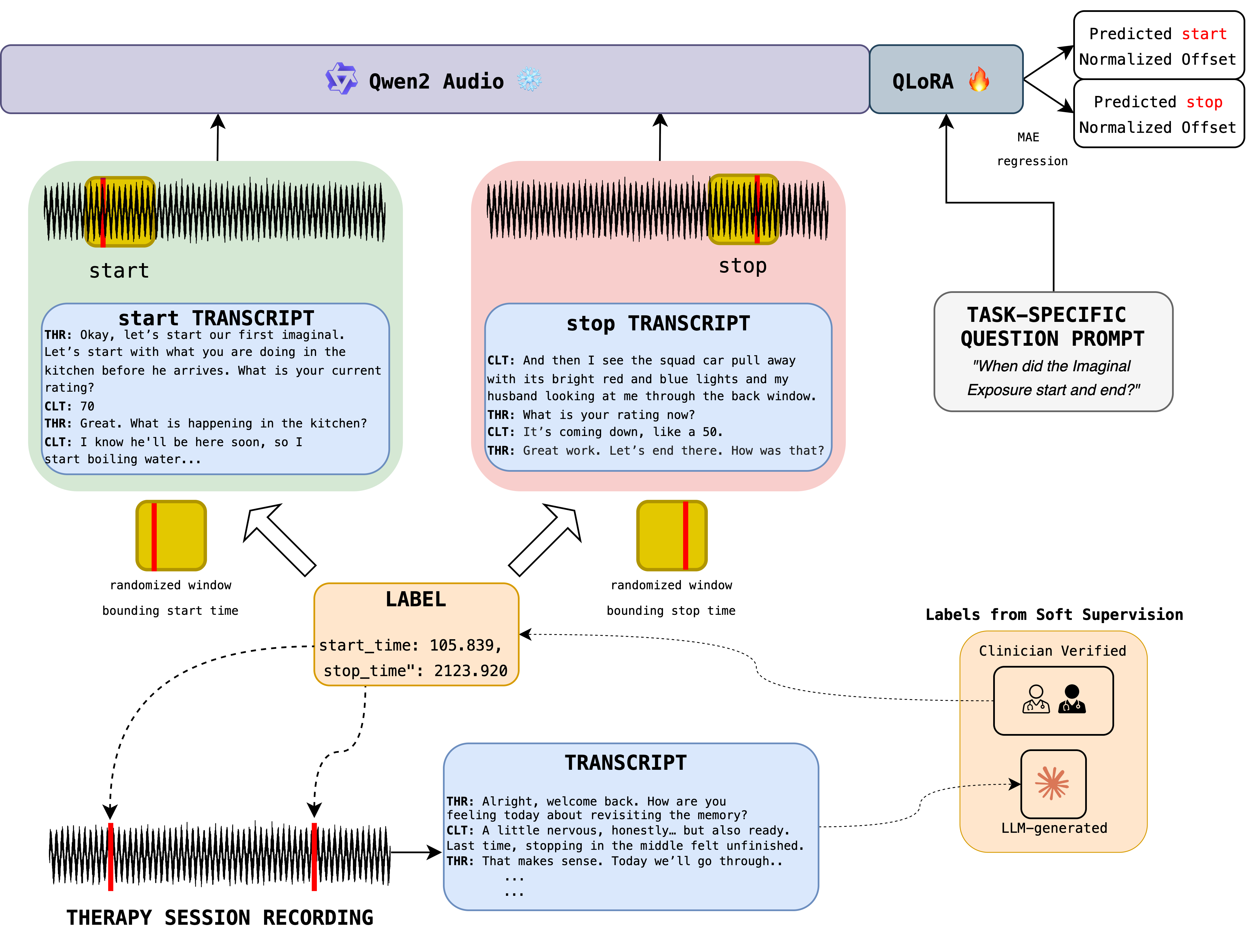}
    \caption{Overview of our fidelity-aligned modeling pipeline. A task-specific prompt guides QLoRA fine-tuning on Qwen2 Audio using 30-120s audio-transcript windows randomly sampled around annotated start/stop points. The model is trained to predict normalized temporal offsets using clinically validated or synthetic labels.}
    \label{fig:block_diagram}
\end{figure}

\begin{table*}[!htbp]
\centering
\caption{Per-phase average, start, and end MAE (in seconds) for timestamp prediction on the test set. We run the experiments on three seeds (42, 78, 123) and report their Mean $\pm$ S.D.}
\label{tab:mae_full_all_seeds_cleaned_1d}
\vspace{0.25cm}
\scriptsize
\begin{tabular}{@{}llccccccccc@{}}
\toprule
\textbf{Window} & \textbf{Config} & \textbf{P1 Avg} & \textbf{P1 Start} & \textbf{P1 End} & \textbf{P2 Avg} & \textbf{P2 Start} & \textbf{P2 End} & \textbf{P3 Avg} & \textbf{P3 Start} & \textbf{P3 End} \\
\midrule
\textbf{30s} & Head Only     & $6.8 \pm 0.1$ & $6.4 \pm 0.2$ & $7.0 \pm 0.3$ & $7.2 \pm 0.2$ & $7.2 \pm 0.2$ & $7.3 \pm 0.3$ & $6.8 \pm 0.2$ & $7.7 \pm 0.6$ & $5.8 \pm 0.4$ \\
             & LoRA (r=2)    & $5.8 \pm 1.7$ & $6.0 \pm 1.5$ & $5.6 \pm 2.2$ & $5.1 \pm 2.3$ & $5.7 \pm 2.7$ & $4.4 \pm 2.3$ & $4.8 \pm 2.0$ & $5.2 \pm 3.0$ & $4.4 \pm 1.9$ \\
             & LoRA (r=4)    & $6.2 \pm 1.3$ & $6.2 \pm 2.0$ & $6.0 \pm 1.2$ & $5.4 \pm 2.0$ & $5.4 \pm 2.2$ & $5.5 \pm 2.0$ & $4.9 \pm 1.5$ & $5.2 \pm 2.0$ & $4.6 \pm 1.4$ \\
             & LoRA (r=8)    & $5.9 \pm 1.4$ & $5.5 \pm 1.5$ & $6.4 \pm 1.4$ & $5.0 \pm 1.8$ & $5.2 \pm 1.8$ & $4.9 \pm 1.8$ & $5.0 \pm 0.5$ & $5.5 \pm 1.8$ & $4.4 \pm 0.8$ \\
\midrule
\textbf{60s} & Head Only     & $12.2 \pm 0.8$ & $11.5 \pm 1.0$ & $12.9 \pm 0.7$ & $13.9 \pm 0.5$ & $13.4 \pm 2.2$ & $14.4 \pm 1.2$ & $13.7 \pm 0.6$ & $15.3 \pm 1.1$ & $12.2 \pm 0.6$ \\
             & LoRA (r=2)    & $11.3 \pm 2.5$ & $11.5 \pm 2.9$ & $11.2 \pm 2.6$ & $12.1 \pm 3.1$ & $12.8 \pm 3.0$ & $11.4 \pm 3.3$ & $12.1 \pm 2.9$ & $13.4 \pm 2.6$ & $10.8 \pm 3.4$ \\
             & LoRA (r=4)    & $11.9 \pm 2.1$ & $11.0 \pm 2.1$ & $12.7 \pm 2.0$ & $10.2 \pm 2.0$ & $10.5 \pm 2.3$ & $9.9 \pm 2.1$ & $11.6 \pm 1.4$ & $11.3 \pm 1.0$ & $11.8 \pm 2.4$ \\
             & LoRA (r=8)    & $9.9 \pm 0.1$ & $9.6 \pm 0.8$ & $10.2 \pm 0.7$ & $9.7 \pm 0.6$ & $10.2 \pm 0.9$ & $9.2 \pm 0.4$ & $10.0 \pm 0.6$ & $9.8 \pm 1.6$ & $10.1 \pm 0.5$ \\
\midrule
\textbf{120s} & Head Only     & $25.2 \pm 2.2$ & $25.0 \pm 2.4$ & $25.4 \pm 2.0$ & $27.4 \pm 0.7$ & $25.1 \pm 1.3$ & $29.6 \pm 2.8$ & $24.4 \pm 2.7$ & $28.1 \pm 3.1$ & $20.8 \pm 2.4$ \\
              & LoRA (r=2)    & $20.7 \pm 2.1$ & $20.8 \pm 2.0$ & $20.5 \pm 2.5$ & $18.0 \pm 2.3$ & $17.5 \pm 2.6$ & $18.5 \pm 2.3$ & $21.4 \pm 1.2$ & $22.6 \pm 0.9$ & $20.2 \pm 1.8$ \\
              & LoRA (r=4)    & $20.7 \pm 1.8$ & $21.4 \pm 2.1$ & $19.9 \pm 2.1$ & $18.8 \pm 2.1$ & $18.3 \pm 3.7$ & $19.2 \pm 1.9$ & $21.0 \pm 1.2$ & $20.3 \pm 2.4$ & $21.7 \pm 0.5$ \\
              & LoRA (r=8)    & $20.1 \pm 1.4$ & $20.3 \pm 1.2$ & $19.8 \pm 1.8$ & $20.7 \pm 1.9$ & $21.7 \pm 2.3$ & $19.6 \pm 1.5$ & $22.3 \pm 1.3$ & $23.5 \pm 3.5$ & $21.2 \pm 1.7$ \\
\bottomrule
\end{tabular}
\end{table*}

\section{METHODOLOGY}
\label{sec:methodology}
\subsection{Problem Formulation}
\vspace{-0.2cm}
We treat fidelity not as a sequence of discrete classifications, but as a continuous temporal regression task. This formulation allows the model to capture the fluid, non-binary transitions typical of therapeutic dialogue, which rigid segmentation often mischaracterizes. We formulate temporal localization of PE therapy fidelity metrics as a regression task. Given a PE therapy recording, transcript, and pre-defined metrics (P1: Orientation; P2: Imaginal exposure; P3: Post-imaginal processing), we predict precise start and end times of each metric instance. To do this, we process focused segments of the session. Let $W_j$ represent an input window $j$, comprising an audio segment $A_j$ and its corresponding transcript excerpt $T_j$. Each window $W_j$ is extracted around a known boundary (start or end) of a specific fidelity metric $q_i$. The true absolute time of this boundary within the full session is $t_{\text{abs}}$. The window $W_j$ has a duration $D_j$ and starts at an absolute time $t_{\text{start},j}$. The true boundary $t_{\text{abs}}$ is therefore located at a normalized offset $o_j = (t_{\text{abs}} - t_{\text{start},j}) / D_j$ within this window, where $o_j \in [0, 1]$. Our model $M$, a fine-tuned audio-language model, takes the audio segment $A_j$, transcript excerpt $T_j$, and a task-specific textual prompt $P(q_i, \text{boundary\_type})$ (where $\text{boundary\_type}$ is \texttt{start} or \texttt{end}) as input. The model is trained to predict the normalized offset $\hat{o}_j = M(A_j, T_j, P(q_i, \text{boundary\_type}))$. The objective is to minimize the error between the predicted normalized offset $\hat{o}_j$ and the true normalized offset $o_j$.

\vspace{-0.2cm}
\subsection{Dataset and Soft-Supervision Labeling}
\vspace{-0.2cm}
Our processed dataset consists of 308 PE therapy sessions, preprocessed as described in Section~\ref{sec:dataset}, and transcribed using Amazon HealthScribe, which provides sentence-level timestamps and speaker attribution. Ground-truth start and stop times for fidelity metrics P1, P2, and P3 were initially generated via soft supervision using Claude Sonnet 3.5, prompted to return timestamps in seconds as illustrated in Table~\ref{tab:llm_annotation}. These LLM-generated annotations were subsequently reviewed and corrected by trained raters to create verified ground-truth labels. All downsampled audio recordings were split at the session level into train/validation/test sets.

\vspace{-0.2cm}
\subsection{Model Architecture and Fine-tuning}
\vspace{-0.2cm}
We fine-tune Qwen2-Audio-7B-Instruct~\cite{chu2024qwen2}, a large audio-language model that processes interleaved audio and text inputs through an audio encoder and a language model backbone. For each trained rater-verified boundary time, we construct focused input windows of fixed durations (30, 60, or 120 sec) containing both audio and aligned transcript segments. To ensure model robustness and enable data augmentation, the true boundary is randomly positioned within each window by normalizing its location to the $[0,1]$ interval and adjusting the window center accordingly. The training target is the normalized offset of the boundary relative to the window start. Each input is prepended with a task-specific prompt, such as: ``The following audio and transcript segment is focused around the START of 'Was Prolonged Exposure done in the session?'. Identify the normalized offset (between 0.0 and 1.0) of this precise START within the given segment.'' We use QLoRA~\cite{dettmers2023qlora} for memory-efficient fine-tuning, applying 4-bit NormalFloat (NF4) quantization with bfloat16 computation, and LoRA~\cite{hu2021lora} to the LLM component with ranks $r \in {2, 4, 8}$, scaling factor $\alpha = 2r$, and dropout of 0.1. A regression head uses the final hidden state of the last non-padding token to predict the normalized offset via LayerNorm, two linear layers with ReLU, and a final Sigmoid. Experiments used three random seeds (42, 78, 123), with results reported as mean $\pm$ standard deviation across these runs.

\section{EXPERIMENTAL SETUP}
\label{sec:experimental_setup}
\vspace{-0.2cm}
We implemented the framework described in Section~\ref{sec:methodology} using Qwen2-Audio-7B-Instruct~\cite{chu2024qwen2} on 308 PE sessions (Section~\ref{sec:dataset}). We also evaluated a ``Head Only'' baseline that trains only the final regression layer to isolate the impact of parameter-efficient fine-tuning on the joint multimodal embeddings. We prioritize this adaptation analysis over unimodal (text-only) ablation because purely text-based approaches are constrained by the discrete granularity of ASR timestamps, which typically lack the continuous resolution required for second-level precision in therapeutic phase transitions. Crucially, therapy phase transitions often occur in the silences or prosodic shifts between words, signals that text-only models inherently discard. Models were fine-tuned with QLoRA using the AdamW optimizer~\cite{loshchilov2019decoupled} (learning rate $1\times10^{-4}$, weight decay 0.01, cosine schedule with 0.1 warmup ratio) for 10 epochs with a batch size of 1 and early stopping (patience=3) based on validation MAE. Predicted offsets were denormalized to absolute timestamps, and mean absolute error (MAE, in seconds) was reported for P1, P2, and P3.

\section{RESULTS AND DISCUSSION}
\label{sec:results_discussion}
\vspace{-0.2cm}
Table~\ref{tab:mae_full_all_seeds_cleaned_1d} presents the mean $\pm$ standard deviation of MAE across three random seeds. Our best-performing configuration, 30-second windows with LoRA rank $r=8$, achieved an average MAE of \textbf{5.3 seconds} compared to ground truth across P1 ($5.9 \pm 1.4$s), P2 ($5.0 \pm 1.8$s), and P3 ($5.0 \pm 0.5$s). Notably, an MAE of 5.3s falls within the natural variance of human cognitive pauses. This suggests the model's localization error is effectively indistinguishable from human rater variability for the purpose of clinical quality control. This level of precision is promising for downstream clinical use, where timestamp-level fidelity tracking is essential. Earlier attempts at feature-level fusion using independent audio and text encoders produced overlapping embeddings across therapeutic phases, limiting their temporal resolution. In contrast, the end-to-end audio-language architecture of Qwen2-Audio-7B enabled learning of joint contextual representations that more effectively captured therapeutic cues. Shorter input windows (30s) consistently resulted in lower MAEs compared to longer windows. For instance, with the ``Head Only'' model, P1 Avg MAE increased from $6.8 \pm 0.1$s (30s) to $12.2 \pm 0.8$s (60s) and $25.2 \pm 2.2$s (120s). This reveals a critical context-granularity trade-off: while larger windows provide more semantic history, they dilute the temporal sharpness required for boundary detection. The 30s window emerges as the optimal balance, retaining sufficient rhetorical context while maximizing edge precision. LoRA fine-tuning generally outperformed the ``Head Only'' baseline, especially at 60s and 120s. At 30s, LoRA $r=2$ and $r=8$ both reduced MAE for P2 Avg to around 5.1s and 5.0s, respectively, compared to 7.2s with ``Head Only''. At 60s, LoRA $r=8$ consistently delivered the best results across phases (e.g., P1 Avg: $9.9 \pm 0.1$s, P2 Avg: $9.7 \pm 0.6$s, P3 Avg: $10.0 \pm 0.6$s) and showed strong stability with low standard deviations. For 120s windows, the trend was less consistent. LoRA $r=2$ yielded the lowest P2 Avg MAE ($18.0 \pm 2.3$s), outperforming both $r=4$ and $r=8$. Similarly, for P3, LoRA $r=4$ slightly outperformed $r=8$. These results suggest that higher LoRA ranks may overfit when the input spans become too broad, and smaller adapters can generalize better in such settings. The optimal LoRA rank varied depending on both window size and phase, highlighting the need for task-specific tuning. LoRA models also tended to show higher variability across seeds compared to ``Head Only'', though configurations like LoRA $r=8$ at 60s achieved both low MAE and strong consistency. Overall, these results reinforce the benefits of combining temporally focused inputs with appropriately tuned model adaptation for timestamp prediction in therapy analysis.

\section{CONCLUSION}
\label{sec:conclusion}
\vspace{-0.2cm}
We introduced a QLoRA-based fine-tuning strategy for Qwen2-Audio to localize therapeutic elements in Prolonged Exposure sessions with high temporal precision (for supervision, $\leq$ 10s is actionable). By leveraging both audio and transcript inputs within focused 30s windows, our approach achieved an MAE of \textbf{5.3s} across rater-verified fidelity phases. Performance is strongly influenced by window size and LoRA rank, with shorter contexts and deeper adaptation yielding the most accurate predictions. These findings support scalable, automated fidelity assessment in PTSD therapy, aiding clinician supervision, training, and quality assurance. By localizing key therapeutic phases automatically, the approach reduces the burden of manual review. Future work will expand to additional fidelity markers and test generalization across diverse therapy settings. Ultimately, this work establishes a blueprint for precise, privacy-preserving compliance tracking that extends beyond PTSD, offering a scalable framework for any high-stakes, protocol-driven conversational domain.




\bibliographystyle{IEEEbib}
\bibliography{refs}

\end{document}